%% file: cir.tex
\def\ROSAT{{\it ROSAT\/\ }}
\def\EXOSAT{{\it EXOSAT\/\ }}
\def\GINGA{{\it GINGA\/\ }}
\def\ASCA{{\it ASCA\/\ }}
\def\RXTE{{\it RXTE\/\ }}

\def\ltsima{$\; \buildrel < \over \sim \;$}
\def\simlt{\lower.5ex\hbox{\ltsima}}
\def\gtsima{$\; \buildrel > \over \sim \;$}
\def\simgt{\lower.5ex\hbox{\gtsima}}

\documentstyle[psfig]{mn}

\title[The iron K complex of Circinus X-1 near zero phase]
{\ASCA observations of the iron K complex of Circinus X-1 near zero phase: spectral evidence for 
partial covering}

\author[W.N. Brandt et~al.]
{\parbox[]{6.5in}{W.N. Brandt,$^1$ A.C. Fabian,$^1$ T. Dotani,$^2$ 
F. Nagase,$^2$ H. Inoue,$^2$ T. Kotani$^3$ and Y. Segawa$^2$}\\
\\
$^1$ Institute of Astronomy, Madingley Road, Cambridge CB3 0HA\\
$^2$ Institute of Space and Astronautical Science, Yoshino-dai, Sagamihara, Kanagawa 229, Japan\\
$^3$ The Institute of Physical and Chemical Research (RIKEN), Hirosawa, 2-1, Wako-shi, Saitama 351-01, Japan\\
}

\begin{document}

\maketitle

\begin{abstract}  
We report on \ASCA energy spectra of Cir X-1 taken near its zero phase on 
1994 August 4--5. The \ASCA SIS detectors allow a much more detailed study of 
the iron K complex than has been possible before. We find that prior to a
sudden upward flux transition the dominant iron K 
feature appears to consist of a large edge from 
neutral or nearly-neutral iron. The depth of the edge corresponds to an absorption
column of $\approx 1.5\times 10^{24}$ cm$^{-2}$, while little absorption
over that expected from the Galaxy is seen at lower X-ray energies. The
differential absorption at high and low X-ray energies combined with 
the iron edge energy are strong evidence that partial covering is a crucial
determinant of the behaviour observed 
from Cir X-1. The continuum spectral variability
observed by \ASCA can also be understood naturally in terms of partial
covering column changes. There is evidence for a relatively weak emission line
from neutral iron with an equivalent width of only about 65 eV. After the
flux transition, the strength of the edge feature is greatly reduced, suggesting
a large reduction in the amount of partial covering. For a large region of 
statistically acceptable chi-squared parameter space, the luminosity
of Cir X-1, after correction for partial covering, need not change during
the transition. We discuss models for the partial covering and suggest
that X-ray scattering by electrons 
may be important. Aspects of the Cir X-1 spectrum
are very similar to those of Seyfert 2 galaxies with Compton-thin tori.
\end{abstract}


\section{Introduction} 

Circinus X-1 (hereafter Cir X-1) is a peculiar and poorly understood 
X-ray binary with an approximately 16.6 day period in which the compact 
object is thought to be a neutron star (Tennant, Fabian \& Shafer 1986). The 
neutron star magnetic field strength, companion star spectral type and 
binary orbital parameters are not well determined. The lack of 
X-ray pulsations and the presence of type 1 X-ray
bursts suggest the neutron star magnetic field strength is less than 
$\sim 10^{11}$ G, and it has been argued that Cir X-1 may be an `atoll' type
X-ray binary with a magnetic field strength of less than $\sim 10^{9}$ G
(Oosterbroek et~al. 1995 and references therein). However, the 
broad band X-ray spectrum of Cir X-1 has been argued to resemble that of a
high magnetic field neutron star rather than that of a low magnetic 
field neutron star (Maisack et~al. 1995; compare with Sunyaev et~al. 1991),
and if Cir X-1 is associated with G$321.9-0.3$ (see Stewart et~al. 1993
and references therein) it would perhaps be difficult to understand 
such a weak magnetic field in such a young neutron star (see figure 1 
of Phinney \& Kulkarni 1994). Type 1 bursting activity from Cir X-1 
appears to be highly intermittent, and to our knowledge no bursts have 
been seen either before or after the \EXOSAT series of 
observations. The optical counterpart to
Cir X-1 has been unambiguously identified by Moneti (1992), but
reddening uncertainties and potential accretion disc contributions
have prevented a precise determination of the spectral type
of the companion star. The orbit has been argued to be highly eccentric due 
to the violent variability seen near zero 
phase (probably associated with the periastron passage of the neutron star), 
although proper orbital parameters are lacking. Given the orbital period 
and claimed space velocity of Cir X-1, it is probable that its orbit was at least 
initially highly eccentric (see figure 10 of Brandt \& Podsiadlowski 1995). 

The iron K spectral features of Cir X-1 and their relevance to understanding 
the X-ray properties of Cir X-1 more generally are also a subject of uncertainty. 
Very strong and potentially broad iron K$\alpha$ emission lines have been 
discussed (e.g. Tennant 1985; Miyamoto \& Kitamoto 1985; Makino 1993; 
Gottwald et~al. 1995; Maisack et~al. 1995), but it is known that complicated
iron K spectral features can be difficult to interpret using
data with low spectral resolution  (e.g. Tennant 1988a; 
Gottwald et~al. 1995). In particular, in Cir X-1 models where a large amount
of partial or total covering is present (e.g. Ikegami 1986; 
section 3.2 of Tennant 1988b; Inoue 1989; 
section 5 of Oosterbroek et~al. 1995), one might expect 
the presence of a strong iron K edge. In this
paper we use the excellent spectral resolution and sensitivity of the 
Japanese/USA Advanced Satellite for Cosmology and Astrophysics 
(hereafter {\it ASCA}; Tanaka, Inoue \& Holt 1994) to probe the iron K
complex of Cir X-1 near one of its zero phase transitions. 
  
Following section 3 of Stewart et~al. (1993) we take the distance to Cir X-1
to be greater than 6.7 kpc. Neither the optical extinction nor the interstellar 
X-ray column to Cir X-1 are precisely known, but the X-ray column is thought to
be about (1--2)$\times 10^{22}$ cm$^{-2}$ (e.g. Predehl \& Schmitt 1995). 
We note that significantly smaller X-ray column values and arguments based 
on them (e.g. section 3 of Stewart et~al. 1991) may have trouble explaining 
the strength of the dust scattering halo of Cir X-1 as shown in figure 7 of 
Predehl \& Schmitt (1995).  

In some of the spectral fitting below we shall implicitly assume 
Morrison \& McCammon (1983) abundances (e.g. when we fit cold absorption and 
partial covering models). To our knowledge there is no evidence which suggests 
that these abundances are incorrect, and scalings to different abundances 
may be made from our results.    

\section{Observations and data reduction} 

Cir X-1 was observed with \ASCA during AO-2 
on 1994 August 4--5 using 
both Solid-state Imaging Spectrometer CCD detectors (SIS0 and SIS1) and 
both Gas Imaging Spectrometer scintillation proportional counters (GIS2 and GIS3). 
The energy resolutions of the SIS and GIS detectors are $\approx 2$ per cent
and $\approx 8$ per cent at the iron K complex. 
The observations started at 20:13:09 UT on 1994 August 4 and ended
at 10:30:45 UT on 1994 August 5. They covered the time of zero phase as
given by equation 1 of Glass (1994) for $N=392$.

The SIS detectors were operated in 1 CCD mode, and the most well-calibrated 
SIS chips were used (chip 1 for SIS0 and chip 3 for SIS1). Bright mode was 
used for high bit rate data and fast mode was used for medium bit rate data. 
We shall use only the bright mode data in this paper, and the telemetry 
limit is 256 count s$^{-1}$ SIS$^{-1}$ for these data. The telemetry is
saturated during the latter part of the observation, and we only analyze
the SIS data for which the telemetry is not saturated. Cir X-1 was placed 
at the nominal 1 CCD mode pointing position. 

The GIS were operated in PH mode, and the bit assignment was changed so as to 
give 6 bits of X and Y positional information and an extra 4 bits of timing 
information (10-6-6-5-0-4 in the notation of section 3.3 of Day et~al. 1995b). 
We thus obtain $64\times 64$ pixel images (i.e. {\tt RAWXBINS} and 
{\tt RAWYBINS} each have the value 64; note however that the GIS field of view
is circular) and 1024 channel energy spectra in the 1--10 keV band. 
The telemetry limit is 128 and 16 count s$^{-1}$ GIS$^{-1}$ for the
high and medium telemetry bit rates, respectively. The telemetry was
completely saturated for the medium bit rate data, and was saturated
for the latter part of the observation even at high bit rate. We shall
use only the high bit rate data in this paper. When GIS photons are lost
due to telemetry saturation, they are lost in a way that leaves
the spectral shape unaffected. Thus GIS data that suffer from telemetry
saturation can be analyzed reliably after suitable corrections (see below). 

We have used the `Revision 1' processed data from Goddard Space
Flight Center (GSFC) for the analysis below (Day et~al. 1995a), and
data reduction was performed using {\sc ftools} and {\sc xselect}
(see Day et~al. 1995b for a description of these packages and
their application to \ASCA data analysis). In this 
paper we focus on the iron K spectral features observed from 
a very bright source, and we have chosen our basic data screening criteria 
after consultation with the GSFC \ASCA help service. The basic screening 
criteria we adopt in this paper are listed in Table 1. We 
have used the {\sc sisclean} and {\sc gisclean} software
in {\sc ftools} as appropriate, and for the SIS detectors we only use event 
grades 0, 2, 3 and 4. 

The {\tt ANG\_DIST} housekeeping parameter is unacceptably large for the 
first $\approx 1500$ s of the observation as expected, but then it remains 
within the criteria of Table 1.

\input table1.tex

\section{Analysis} 

We first extracted images for Cir X-1 in each detector and examined 
them. At low energies the profile of Cir X-1 has extent 
over that expected for a point source due to its dust scattering halo (see
Predehl \& Schmitt 1995 for a detailed study of the dust scattering halo 
using higher spatial resolution data from the \ROSAT PSPC). This dust
halo will seriously hinder X-ray imaging searches for jet emission. At 
higher energies (above $\approx 5$ keV), the profile of Cir X-1 is
consistent with that of a point source (this is as expected since dust
scattering halo intensities drop strongly with increasing energy; see 
section 3.6 of Predehl \& Schmitt 1995). There are no other strong
X-ray sources in the fields of view. 
However, the GIS detectors also show weak apparent emission to the south 
of Cir X-1 at their edges (this emission is not from the calibration sources). 
This `shoulder-like' emission is part of the point spread function and is common in 
observations of bright sources (K. Ebisawa, private communication). We are 
confident that this emission originates from Cir X-1 and is not from another 
X-ray source because its variability tracks the variability from Cir X-1 
described below (with reduced absolute amplitude).       

We use circular source cells centered on Cir X-1 with radii of 
$\approx 4.5$ arcmin and $\approx 9$ arcmin for SIS and GIS source count
extraction, respectively. The choice of background regions is not 
straightforward for the SIS detectors. The brightness of Cir X-1
requires us to use small circular regions of radius $\approx 1.2$ arcmin
near the northeastern (SIS0) and southeastern (SIS1) chip corners. Even
with these regions, there is still some contribution from Cir X-1 to 
the background cells. For the GIS detectors we use circular background cells 
with radii of $\approx 2.0$ arcmin that are essentially free of emission from 
Cir X-1. We have found that different background regions have very little 
effect on the results of the analysis presented below.  

\subsection{Variability near zero phase}

\subsubsection{Count rate variability}

\begin{figure*}
\centerline{\psfig{figure=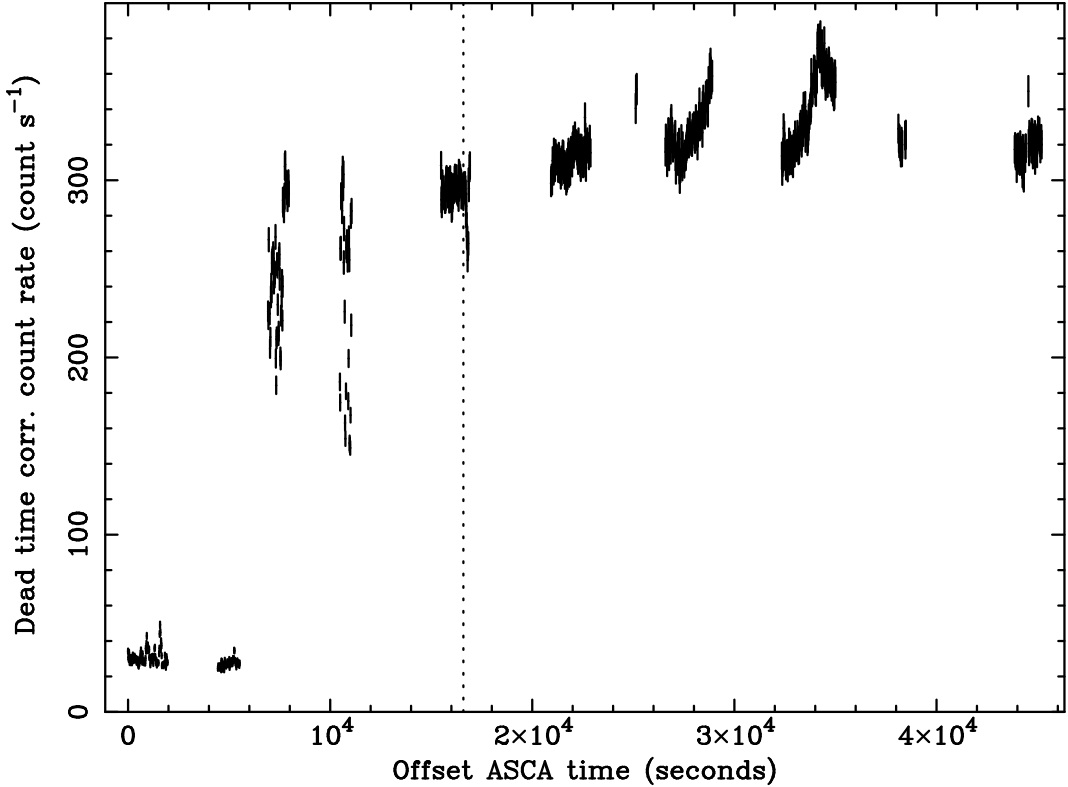,width=0.9\textwidth,angle=270}}
\caption{Screened \ASCA GIS2 light curve of Cir X-1 after correction for detector dead time 
(see the text for the correction method used). 
The light curve is for the energy band 1.0--10.0 keV (GIS channels 85--849).  
The abscissa is in seconds after 20:37:12 UT on 1994 August 4, and the data point bin size is 16 seconds.    
The vertical dotted line shows the predicted time of zero phase from equation 1 of Glass (1994) with $N=392$.
We shall refer to the first 6000 s of data as the `low count rate state' and the data after this time
as the `high count rate state.' There is very rapid and large amplitude variability 
between 6000--12000 s, and this variability causes the rough appearance of the data in 
this time interval.}
\end{figure*}

\begin{figure*}
\centerline{\psfig{figure=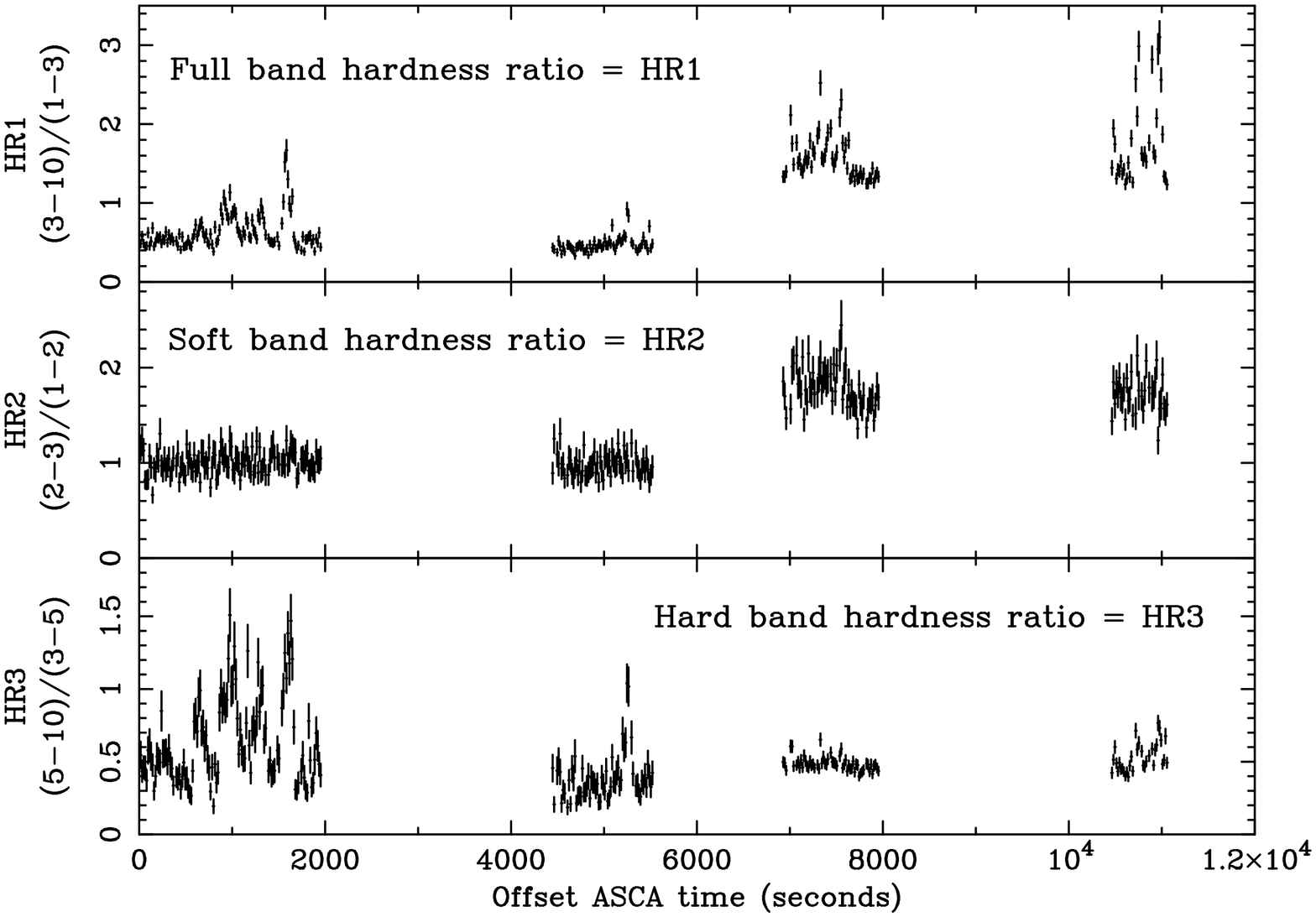,width=0.9\textwidth,angle=270}}
\caption{\ASCA GIS2 hardness ratios of Cir X-1 as a function of time. 
The ordinate labels denote energy ranges in keV. 
The abscissa offset is the same as that for Figure 1, and we have only shown 
the start of the observation where the strongest spectral variability is observed. 
The data point bin size is 16 seconds. Note that HR3 can vary strongly while
HR2 remains constant.}
\end{figure*}

Source and background light curves were extracted for all instruments. We
have corrected the GIS light curves for dead time effects by
multiplying each point by the (time varying) factor 
1/(1$-${\tt G2\_DEADT}) for GIS2 and 
1/(1$-${\tt G3\_DEADT}) for GIS3.
The {\tt G2\_DEADT} and {\tt G3\_DEADT} parameters are obtained from the \ASCA 
househeeping file, and they include the effects of both instrumental and
telemetry dead time. When we quote GIS count rates 
below we implicitly include the correction
for detector dead time. We found good general agreement between the shapes of the source 
light curves. There was also good general agreement between the shapes of these
light curves and the shapes of the {\tt G2\_L1} and {\tt G3\_L1} light curves.
There is no evidence for strong background flaring, even when the screening
criteria shown in Table 1 are neglected (variability in the SIS background
light curves can be attributed to the small contamination from Cir X-1 itself). 

In Figure 1 we show the screened GIS2 light curve of Cir X-1 in the 1--10 keV band. 
Cir X-1 is in a relatively low count rate state ($\approx 30$ count s$^{-1}$) 
for about 6000 s, and it then makes an abrupt transition to a state 
where the mean GIS count rate is a factor of $\sim 10$ times higher.
Given the detector position of Cir X-1, it has a GIS count
rate of about 0.05 times that of the Crab during the low count rate
state and a GIS count rate of about 0.55 times that of the Crab during
the high count rate state. The high count rate state 
shows strong and rapid variability for its
first $\approx 6000$ s, but Cir X-1 then settles into a calmer mode
of behaviour. The Eddington limit of Cir X-1 for hydrogen-poor material is 
thought to correspond to a flux of about $2.4\times 10^{-8}$ erg cm$^{-2}$ s$^{-1}$
(see section 2 of Stewart et~al. 1991 but be wary of partial covering 
effects as described below), and in the high count rate state the flux
after correction for only Galactic absorption is comparable to the 
Eddington limit flux (see Table 2 for numerical flux values). 

The basic shape of the zero phase light curve has been seen to change 
greatly over the years (see the discussion in section 3 of Tennant 1988b
and Whitlock \& Tyler 1994). Our observations 
do not show evidence for basic shape changes from the \GINGA
all-sky monitor and pointed observations reported by
Tsunemi et~al. (1989) and Makino (1993), although the rise
after zero phase may be to a somewhat lower level
\footnote{We also made \ASCA observations of Cir X-1 on 
1994 August 26. These observations are about 5 days past zero phase
according to equation 1 of Glass (1994) for $N=393$, and the
observed mean GIS2 count rate is $\approx 350$ count s$^{-1}$ with 
$\approx 10$ per cent variability. Comparison of this fact with
Figure 1,
figure 1 of Tennant (1988a),  
figure 8 of Tsunemi et~al. (1989) and
\RXTE all-sky monitor data 
suggests that the `decay' of the flux from Cir X-1 after zero phase 
may be progressively occuring less rapidly. Further 
observations to search for such trends are needed.}.  
We have verified this via further examination of the unpublished
\GINGA all-sky monitor data which show that Cir X-1 does not appear to
significantly exceed 1.3 Crab in the 1--20 keV band during 1987--1992
(note that the `1.8' and `0.6' in the caption for figure 8 of 
Tsunemi et~al. 1989 are incorrectly transposed; see figure 7 of
Tsunemi et~al. 1989 and H. Tsumeni, private communication). 
\RXTE all-sky monitor observations from 1996 show flux dips before
zero phase as well as flux rises up to a remarkable 2--3 Crab shortly after
(within about 0.2--0.6 days) zero phase (see Shirey et~al. 1996). It 
is hard to determine
whether the data from the \GINGA all-sky monitor or the \RXTE all-sky
monitor are more germane for comparison with our 1994 \ASCA data, 
although we do not see evidence for a count rate rise as dramatic as
is now seen by {\it RXTE\/}. In any case, we note that the observed 
variability supports the discussion in section 3 of Glass (1994) regarding 
ephemerides. 

The high count rate state variability at 28000--36000 s in Figure 1 is of 
concern since the two rises appear to have similar shapes. We have
manually examined the {\tt ANG\_DIST} parameter during this time,
and while small fluctuations of $\approx 9$ arcsec or less are
present, they do not appear to be able to explain the variability. 
This variability is also seen in the 
{\tt G2\_L1},
{\tt G2\_LDHIT},
{\tt G3\_L1} and
{\tt G3\_LDHIT}
housekeeping parameters, and the 
{\tt G2\_H0},
{\tt G2\_H2},
{\tt G3\_H0} and
{\tt G3\_H2}
housekeeping parameters appear to be reasonably behaved during
this time. The cut-off rigidity is well above 4 GeV c$^{-1}$ 
during this time.  

\subsubsection{Spectral variability}

In Figure 2 we show GIS2 hardness ratios as a function of time. 
We define HR1 to be the ratio of the count rate in GIS2 channels 
256--849 to that in channels 85--255 (these energy ranges approximately 
correspond to 3.0--10.0 keV and 1.0--3.0 keV). 
We define HR2 to be the ratio of the count rate in GIS2 channels 
171--255 to that in channels 85--170 (these energy ranges approximately 
correspond to 2.0--3.0 keV and 1.0--2.0 keV). 
We define HR3 to be the ratio of the count rate in GIS2 channels 
426--849 to that in channels 256--425 (these energy ranges approximately 
correspond to 5.0--10.0 keV and 3.0--5.0 keV). 
Note that we have reduced the length of the abscissa of Figure 2
relative to that of Figure 1 to only show the period of strongest spectral 
variability. 

During the low count rate state (the first 6000 s of Figure 2) Cir X-1 shows 
strong HR1 and HR3 variability but much less HR2 variability. The HR1 and HR3 
variability often has a flare-like character. The 
HR2 data during the low count rate state can be fit acceptably with a constant 
value of 0.969 ($\chi^2_\nu=0.97$ for 181 degrees of freedom), although there
do appear to be some small systematic residuals. In light of
this result we have examined the 1.0--3.0 keV light curve, 
and during the low count rate state the variability in this band is much 
less than that at higher energies. While the data are not
strictly consistent with a constant model
($\chi^2_\nu=1.17$ for 181 degrees of freedom), no systematic count rate
fluctuations by more than 20 per cent from the mean are apparent. In 
addition, the flare-like behaviour seen at higher energy is not obviously 
visible below 3 keV.   

All three hardness ratios are seen to change significantly after the
transition from the low count rate state to the high count rate 
state, although the change at high energies appears to be less
than that at low energies. 

After 12000 s (not shown in Figure 2), much less hardness ratio
variability is seen. In particular, there is no apparent spectral
variability corresponding to the count rate variability seen
in Figure 1 between 28000--36000 s. Statistical analysis shows that 
there is a slow general decline of at least HR1 and HR3 with
time. HR1 declines from 1.33 to 1.22, and HR3 declines from 
0.46 to 0.44. 

\subsection{Spectra near zero phase}

Our \ASCA observation provides the highest spectral resolution view of Cir X-1
yet possible and allows a sensitive probe of its iron K spectral features. To
model the continuum underlying these features, we have used a two blackbody
model. As we shall show below, this model fits the continuum in the \ASCA band
well even if the fitted blackbody parameters may not have precise translations
into physical quantities. This continuum model was also found
to fit the broader band ($\approx$2--25 keV) and very high statistical quality 
\GINGA data the best (Makino 1993). We have verified that our conclusions regarding
the iron K spectral features are not materially altered if we use different
continuum models such as Comptonized blackbodies (this is due to the good
spectral resolution and good statistics). In addition, if we ignore
the data below 3 keV and just use a power law to fit the 4--10 keV data our
results are not qualitatively changed. In all fits below we include the
effects of interstellar absorption. In our fits with partial covering
we, of course, take the interstellar absorption to be unaffected by the
partial covering. 

Photon pile-up effects are a concern for \ASCA SIS observations of bright
sources (see Ebisawa et~al. 1996 for a detailed discussion of this issue). 
Cir X-1 has a mean count rate during the low count rate state of 
$\approx 30$ count s$^{-1}$ SIS$^{-1}$, and photon pile-up effects are unlikely
to compromise the analysis below. We have used ring-shaped extraction regions
to verify that our results are not strongly influenced by photon pile-up
effects. We note that the pile-up spectrum is smooth (see figure 11 of
Ebisawa et~al. 1996) and cannot falsely imprint the sharp spectral features
we describe below. 

We use  the SIS redistribution matrix files (rmf) from 1994 Nov 9 and 
the GIS rmf from 1995 March 6. We generate our ancillary response files (arf) 
using the {\sc ascaarf} software. We have used the {\sc deadtime} program 
in {\sc ftools} to correct for dead time effects. We model the X-ray spectra 
of Cir X-1 using the X-ray spectral models in the {\sc xspec} spectral fitting 
package (Shafer et~al. 1991). 

\subsubsection{Low count rate state spectral fitting}

\begin{figure*}
\centerline{\psfig{figure=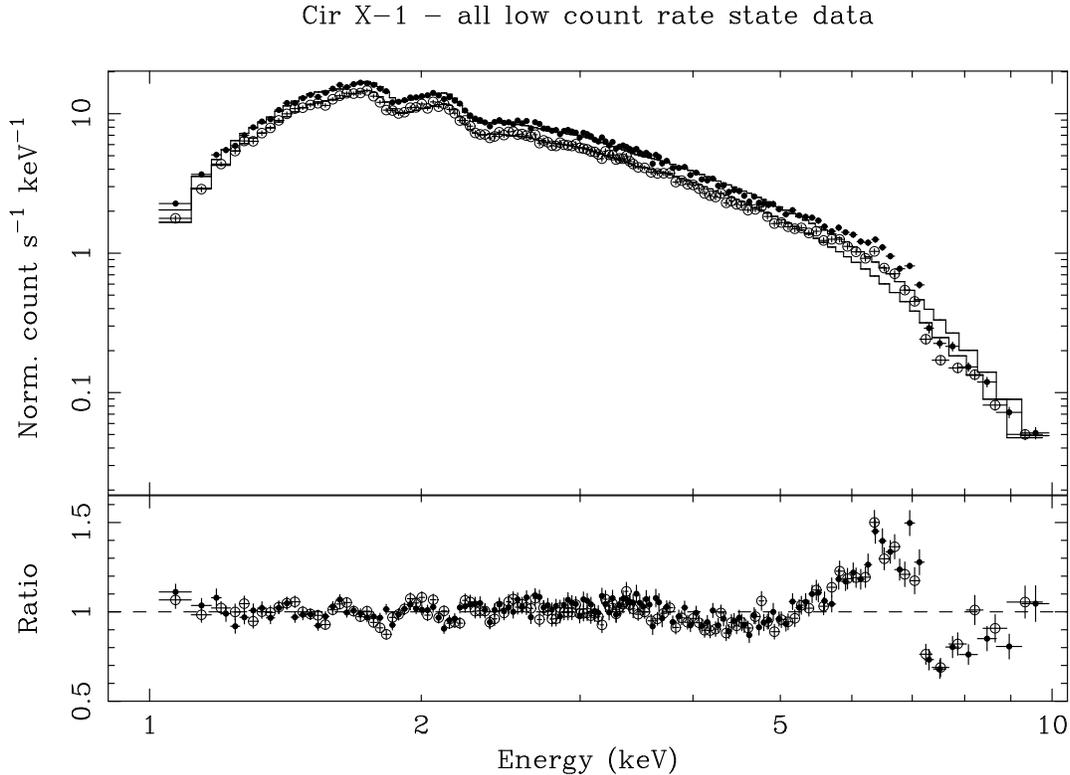,width=0.9\textwidth,angle=270}}
\caption{Average SIS0 (solid circles) and SIS1 (open circles) spectra for the low count rate state
data. The best fitting absorbed two blackbody model is shown along with the data-to-model
ratio. Note the sharp drop in the data-to-model ratio just above 7 keV. The SIS detector
resolution at the iron complex is about 140 eV, and the shown data points in this
energy range are essentially statistically independent. The sharp drop
can be interpreted as an iron edge from partial covering.}
\end{figure*}

\vskip 0.1 cm
\noindent {\it Average spectral properties}
\vskip 0.1 cm

\noindent In order to obtain a `first look' at the spectrum in the low count rate state, we
have binned spectra for all four detectors using all of the low count rate state 
data. We group these spectra so that there are at least 20 photons per spectral
data point (we shall implicitly adopt this grouping for all spectra below unless stated 
otherwise). Figure 2 shows that there is significant spectral variability within the 
low count rate state, and thus these spectra must be regarded to be
crude and only meaningful in a time-averaged sense. However, the excellent statistics 
afforded by this first look give useful insight. We note that an unlikely conspiracy 
would be required for spectral variability effects to produce sharp spectral features at 
the energies of known strong atomic features. 

In Figure 3 we show average SIS0 and SIS1 spectra for the entire low count rate
state. The two spectra agree well in shape, and the GIS2 and GIS3 spectra 
are consistent with those in Figure 3 (they are not displayed only for reasons of clarity). We 
have used the two blackbody model with simple cold absorption (see above) to fit these 
data, and we show the ratio of the data to the best fitting model. In this fitting the 
blackbody normalizations are all taken to be free parameters, but the other spectral 
parameters are tied together (we adopt this fitting prescription for all fits below
unless stated otherwise). The most noticeable feature in the ratio plot is the large and 
sharp drop just above 7 keV. A natural interpretation for this drop is that it is due to
an absorption edge from a large column of neutral or nearly-neutral iron atoms. The 
broad 5.3--7.1 keV excess and smaller 3.8--5.3 keV deficit may be understood 
as statistical artefacts from the fit 
trying to compensate for the presence of the large edge (fits which include the edge reduce 
the amplitudes of these features to 
where they are insignificant, see below). A large edge from iron would be 
expected in models for Cir X-1 where a large amount of obscuring matter lies between 
the X-ray generation region and Earth, and one would not be naturally 
expected in other models. The iron structure could perhaps also be interpreted
as broad iron line emission from matter moving with relativistic speed, and we shall 
critically examine this possibility below. There is 
also some secondary residual structure between 6--7 keV, although this 
structure is not as strong as that just above 7 keV. There are no strong residuals 
in the 1--3 keV band.  

\vskip 0.1 cm
\noindent {\it `Calm' set of data}
\vskip 0.1 cm

\input table2.tex

\begin{figure*}
\centerline{\psfig{figure=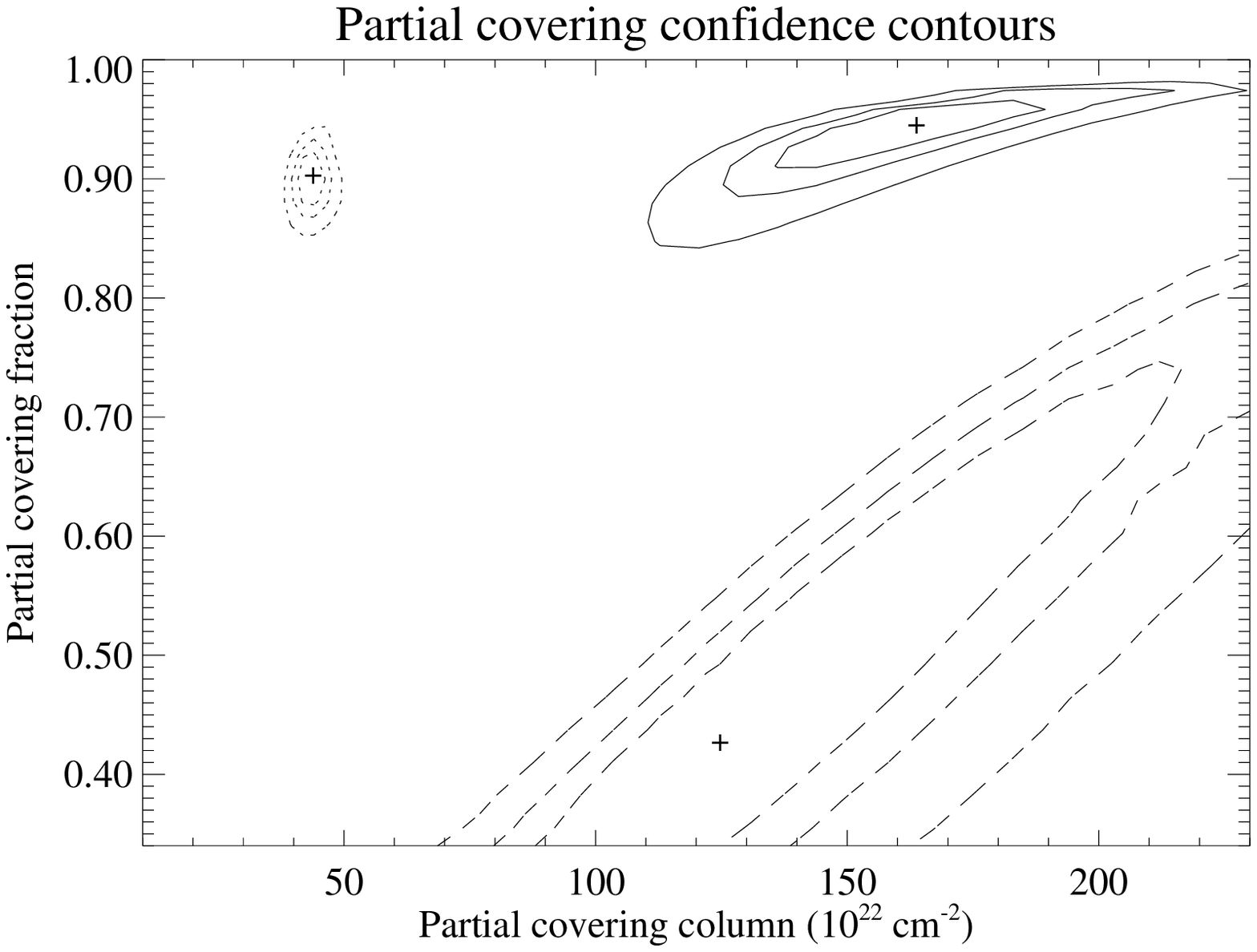,width=0.9\textwidth}}
\caption{Contours of the partial covering model for 
the `calm' low count rate state data (solid curves near upper right), 
the low count rate state data with HR3 greater than 0.9 (dotted curves near upper left) and 
the high count rate state data (dashed curves). 
The partial covering is taken to affect both blackbodies (see the text). Contour 
levels for each set are for
$\Delta\chi^2=2.30$ (68.3 per cent confidence for two parameters of interest),
$\Delta\chi^2=4.61$ (90.0 per cent confidence for two parameters of interest) and
$\Delta\chi^2=9.24$ (90.0 per cent confidence for five parameters of interest).
Crosses show the best fit values.
The low count rate state contours have been made using data from all four
\ASCA detectors, while the high count rate state countours have been made only
using data from the GIS detectors. The partial covering luminosity correction
increases towards the upper part of this diagram, especially for partial covering 
fractions near unity.}
\end{figure*}

\noindent In order to probe the spectral properties of the low count rate state in a more 
precise manner, we have carefully selected a 450 s interval of data where there 
is relatively little flux and spectral variability. This interval begins 
at 21:55:31 UT. We have binned spectra for all four detectors and fit these
spectra using the two blackbody model with simple cold absorption. We derive
the results shown in column 1 of Table 2. 
This model can be ruled out with greater than 94 per cent confidence, 
and similar systematic residuals to those shown in Figure 3 are present.

The depth of the putative edge seen in the residuals and in Figure 3 is 
suggestive of intrinsic absorption by a column 
of $\sim 10^{24}$ cm$^{-2}$ even though we do not see significant 
absorption over the expected Galactic value at lower X-ray energies. This 
large differential column could either be explained by partial covering from
thick matter or by absorption from ionized matter. Due to the fact that the 
putative edge energy corresponds to that of neutral or nearly-neutral iron, 
partial covering seems more plausible than warm absorption. If we include
partial covering of both blackbodies, we derive the results shown in
column 2 of Table 2. 
The fit is statistically very good. In Figure 4 we show partial covering 
chi-squared contours for this fit as the solid set of contours. Note 
that a very large partial covering fraction is required 
by these data and that as a result the partial covering corrected
isotropic luminosity for the low count rate state is of order the Eddington 
luminosity (it is difficult to be more specific than this because
the observed X-ray bursts could have been affected at some level
by partial covering too).  

Narrow emission lines from iron have been seen in X-ray binaries
(e.g. Ebisawa et~al. 1996), and Figure 3 suggests the presence of
an emission line from neutral or nearly-neutral iron. We have thus
added a 6.40 keV (see House 1969)
Gaussian emission line with $\sigma=10$ eV to our fit. This addition
gives $\chi^2=924.3$ for 929 degrees of freedom, a significant
improvement (the other fit parameters are not changed significantly
by the addition of the line). Adding narrow lines at other energies
(e.g. 4.0 keV, 6.1 keV, 6.6 keV) does not improve the fit
significantly. The line flux is 
$(8.9^{+8.9}_{-5.6})\times 10^{-3}$ photons cm$^{-2}$ s$^{-1}$, and the 
line equivalent width is $65^{+66}_{-41}$ eV (these errors are for
$\Delta\chi^2=2.71$). The equivalent width we obtain from fitting
the data shown in Figure 3 is comparable to this value.
While there may also be small 6.7--6.9 keV lines
from ionized iron, we are not able to formally prove the existence of such 
lines. The other detectors do not have $\approx 6.9$ keV residuals
as strong as the SIS0 $\approx 6.9$ keV point shown in Figure 3. 

We have also considered the possibility that only the higher
temperature blackbody component suffers from partial covering.
While figure 4 of Inoue (1989) appears to be at odds with this
possibility, we note that HR2 does not vary during the low count rate 
state while HR3 does. The absence of HR2 variability, the absence
of large 1.0--3.0 keV count rate variability (see above) and
the presence of HR3 variability could be suggestive of partial
covering interposed between the higher and lower temperature 
blackbody emission regions (but see our further discussion of this
issue below in the context of partial covering column changes). 
From our fitting we derive the results shown in column 3 of Table 2. 
This is also a statistically good fit and thus from model fitting alone 
we cannot clearly determine whether the partial covering affects both 
blackbodies or only the higher temperature one. If we make partial covering 
contours such as those shown in Figure 4, they look very similar to the 
solid ones already shown. If we use the low and high temperature
blackbody normalizations and a distance of 6.7 kpc to derive crude 
blackbody radii, the higher temperature 
blackbody has a radius of $\simlt 36$ km and the
lower temperature blackbody has a radius of $\simlt 52$ km. Thus
in this scenario the matter doing the partial covering would 
be roughly $\sim 40$ km from the neutron star. A very large
density of $\simgt 10^{22}$ cm$^{-3}$ would be required for iron
to be nearly neutral, as implied by the energy of the iron 
edge (see Ross 1978 and Kallman \& McCray 1982). This density 
is remarkably high, being $\sim 1400$ times larger than the Eddington 
number density for accretion onto a 1.4 M$_\odot$ neutron star 
($n_{\rm Edd}=L_{\rm Edd}/4\pi r_{\rm g}^2c^3m_{\rm p}=10^{19}(M/M_\odot)^{-1}$ cm$^{-3}$
where $r_{\rm g}=GM/c^2$; see appendix C of Phinney 1983 and compare this density 
with section 5.6 of Frank, King \& Raine 1992). The presence of such a 
dense partial coverer would be difficult to understand. Thus
partial covering of just one blackbody, while statistically acceptable,
does not appear to be physically plausible. The lack of strong Doppler
blurring of the edge may also suggest that the partial coverer is 
significantly further than $\sim 40$ km from the neutron star. If the partial 
covering affects both blackbodies, we derive a radius for the 
lower temperature blackbody of $\sim 170$ km, and the partial coverer 
could be far outside this radius. Thus scenarios in which the 
partial covering affects both blackbodies significantly alleviate the 
density contrast problem. 

Now we consider the possibility that the iron K residuals are primarily due 
to a broad iron emission line from matter moving with relativistic speed
near the neutron star (e.g. Brandt \& Matt 1994 calculate the expected 
iron K$\alpha$ line properties from matter in an ionized accretion disc 
around a neutron star). The blue horns of such lines can sometimes drop 
quite sharply at their high energy ends, and this could perhaps produce the
sharp drop shown in Figure 3. We have fit the {\sc xspec} `diskline' model
to the iron K spectral features. This model includes Doppler, transverse Doppler 
and gravitational redshift effects but does not include light bending
(Fabian et~al. 1989 describe the physics used in the `diskline' model but note 
that not all of the physics discussed in Fabian et~al. 1989 is actually 
implemented in the model). We constrain the line rest energy to be in the
range 6.40--6.97 keV 
(this is the iron K line energy range from House 1969 and Ross 1978). 
We fix the disc inner radius to be $6r_{\rm g}$ and constrain
the disc outer radius to be larger than $7r_{\rm g}$. We have found that the 
statistically best fits are obtained when we use the special `10' disc
emissivity law appropriate for a black 
hole disc (see Shafer et~al. 1991 and the {\sc xspec} `diskline' code).
In this fit the line energy rises to 6.97 keV and the inclination drops 
to 0 degrees. The best fitting disc outer radius is $110r_{\rm g}$, but this 
parameter is poorly constrained by these data. The line equivalent width
is 440 eV. This equivalent width, while large, is perhaps not
impossible for an ionized accretion disc (e.g. Matt, Fabian \& Ross 1993).
The fit gives $\chi^2=943.4$ for 928 degrees of freedom, and is 
thus statistically acceptable. From examination of the best fitting line 
model, it is apparent that the high energy drop of the blue horn of the 
line mimics the effect of an edge, as discussed above. We note that while a 
relativistic line cannot be formally falsified with these data, 
partial covering provides a slightly smaller $\chi^2$ with
fewer free parameters ($\Delta\chi^2=11.8$ with 2 fewer parameters). In 
particular, the drop just above 7 keV is uniquely and 
naturally specified in the partial covering model by 
the edge energy of cold iron. In addition, aspects of the observed spectral
variability and low-to-high count rate state transition can be best 
explained in terms of partial covering (see below). 

In light of the results of this section, it appears likely that some 
earlier and lower spectral resolution observations which modeled the 
iron K complex of Cir X-1 with only a line feature were not entirely
correct. 

\vskip 0.1 cm
\noindent {\it Other sets of data}
\vskip 0.1 cm

\noindent We have also fit other sets of data to investigate the extent to
which our discussion above is generally applicable to the low count rate 
state. The first 500 s of data in Figure 1 show relatively little
flux and spectral variability, and the spectral residuals from these 
data are consistent with those discussed 
above. If we fit a partial covering model to 
these data, we obtain a set of contours that is very similar to the
solid set shown in Figure 4. The partial covering fraction is again
required to be greater than 80 per cent with very high statistical
significance. 

\begin{figure*}
\centerline{\psfig{figure=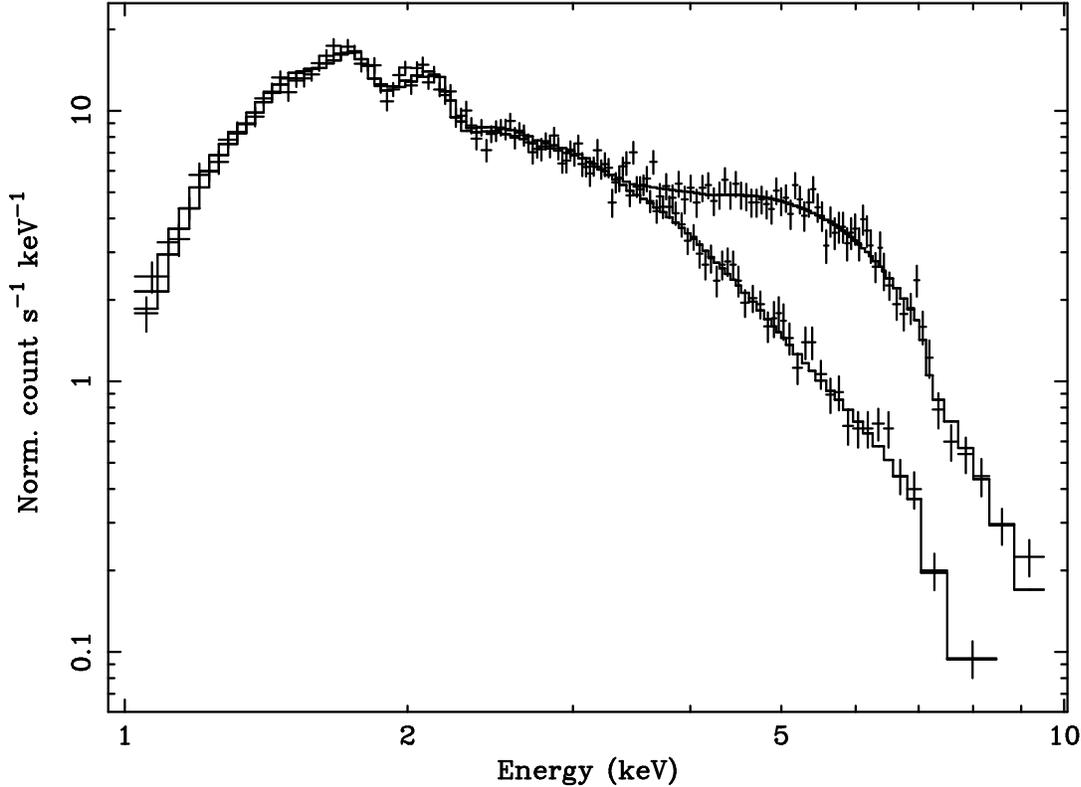,width=0.9\textwidth,angle=270}}
\caption{SIS0 spectra extracted from the `calm' set of data (lower spectrum) 
and a set of data which has HR3 greater than 0.9 (upper spectrum). Note that the 
spectra agree very well up to about 3 keV but then differ strongly at higher 
energies. The data sets have been fitted with double blackbody models which
include partial covering (see the text). A partial-covering column change
can explain the difference between these two spectra (see the text).}
\end{figure*}

Unfortunately, there are no other contiguous sets of low count rate state 
data during which both HR1 and HR3 remain well-behaved for 200 s or more 
(200 s is about the time required to obtain a high quality spectrum during the 
low count rate state). In order to proceed further, we are forced to mollify 
our rigorous prescription of only looking at contiguous data sets with
minimal spectral variability. We have selected
a set of low count rate state data for analysis during which HR3 is larger
than 0.9. The data used were chosen from that taken between 800--1700 s in 
Figure 2, and the total integration time was 280 s.   
Our choice of these data allows us to examine a significantly different HR3 
range from that examined so far. In Figure 5 we illustrate the difference 
between a spectrum extracted from these data and one from the `calm' data 
described above. As one would expect from Figure 2 and its
accompanying discussion, the spectra agree well below 3 keV but are 
very different at higher energies. If, as argued above, the partial covering 
affects both blackbodies, the excellent agreement below 3 keV rules out
any significant changes in the partial-covering fraction between the
two data sets. However, the agreement does {\it not\/} rule out changes in 
the partial-covering column density between the two data sets as long
as the column remains so high that it always entirely blocks the
$<3$ keV X-rays that strike it. In fact, the qualitative behaviour shown in 
Figure 5 is exactly what one would expect from a thick partial covering column
that changes with time (this can be understood by examining {\sc xspec}
partial covering models with partial covering columns between
$5\times 10^{23}$ cm$^{-2}$ and $10^{25}$ cm$^{-2}$). If we fit the
spectra from all four detectors with only a two blackbody model, the fit is very 
poor and can be rejected with greater than 99 per cent confidence. If we add 
partial covering, we obtain a good fit as shown in column 4 of Table 2.  
We note that the low energy spectral parameters ($N_{\rm H}$, $kT_1$ and $A_1$) 
and the partial covering fraction agree well with what was derived for the `calm' 
set of low count rate state data. While the $kT_2$ and $A_2$ contours
do not quite overlap with those derived for the `calm' data,
they are not strongly different and small discrepancies may be explained
by the fact that we have fit a set of data which includes HR3
spectral variability. However, the most noticable change from the
`calm' data set is the strong change in fitted partial covering
column. This can be clearly seen by comparison of the dotted and solid 
contours in Figure 4. The dotted contours are smaller than the
solid contours because lower energy photons are now able to penetrate
the column and thereby better constrain the fit. 
                                                         
We have also examined other sets of data with HR3 between 0.6--0.9. 
When we make partial covering contours for these data sets, the contours 
lie between the dotted and solid sets of contours in Figure 4. These data 
sets all have partial covering fractions larger than 0.8 with high 
statistical significance. When we plot them in Figure 5 they lie
in between the two curves shown (i.e. the spectrum `unzips' from
high to low energy as the partial covering column decreases). 

Physically, a large change in the partial covering column without some 
change in the partial covering fraction would appear somewhat surprising. 
Below we shall discuss ways in which the apparent partial covering 
column might change without an accompanying change in the apparent 
partial covering fraction. We shall also discuss how one might
observe apparent partial covering effects from an object as small as a 
neutron star and its inner accretion disc.  

\subsubsection{High count rate state spectral fitting}

\begin{figure*}
\centerline{\psfig{figure=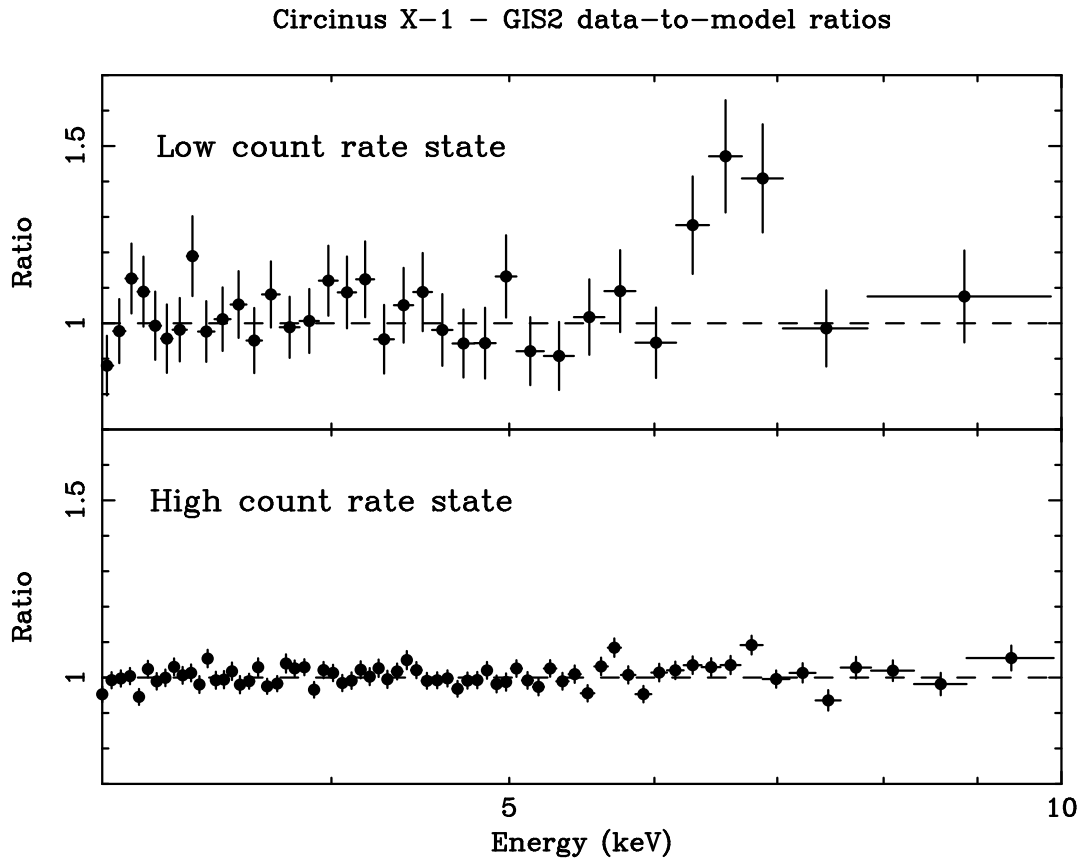,width=0.9\textwidth,angle=270}}
\caption{GIS2 data-to-model ratio plots for the low count rate state data 
(upper panel) and the high count rate state data (lower panel). Note that
the abscissae and ordinates of both panels have the same scales, and that
the low count rate state data have much stronger iron K complex residual
structure than the high count rate state data. The spectral resolution of
the data in this figure is about 4 times less than that in Figure 3.}
\end{figure*}

Due to telemetry saturation during the high count rate state, we have only
been able to perform reliable analyses using the GIS detectors. To begin, we
have extracted GIS spectra using only data taken within the 20000--24000 s 
interval of Figure 1. We group these spectra so that there are at least
40 photons per spectral data point. During this interval there is no strong
spectral variability and the flux variability is rather small as well. If we
fit the two blackbody model with simple cold absorption to these data, we obtain
the results shown in column 5 of Table 2. 
While this fit is statistically unacceptable and can be rejected with greater 
than 98 per cent confidence, examination of the data-to-model ratio reveals
that fractional deviations from the model are much smaller in the high
count rate state than the low count rate state. This important result
is illustrated in Figure 6 where we compare data-to-model ratios for the
high and low count rate states. The difference in statistical fit quality is 
thus primarily due to the much larger number of photons available for 
spectrum construction during the high count rate state. Scrutiny of the
high count rate state residuals requires caution due to the fact that
any systematic residuals are so small that calibration uncertainties may 
have non-negligible effects on their interpretation (see appendices A and B
of Ebisawa et~al. 1996). Careful examination suggests that there may be a 
weak edge-like residual present around 7 keV, and this seems at least plausible 
in light of our results from the previous section. If we add partial covering 
to our model as per the previous section, we obtain the results shown in
column 6 of Table 2. 
The addition of partial covering gives a significant reduction of $\chi^2$, 
although the fit is still not formally acceptable. 
There are no systematic trends in the residuals that are obviously suggestive
of spectral features, and we suspect that calibration uncertainties are
contributing significantly to the $\chi^2$ value due to the very good
statistics. If we include 2 per cent calibration uncertainties when we
bin our spectra (see Tashiro et~al. 1995 for a discussion of GIS 
calibration uncertainties), we are able to obtain statistically acceptable
$\chi^2$ values. In Figure 4 we show partial covering confidence contours 
as the dashed set of contours. The high count rate state contours in this diagram
are larger due to the fact that the strength of the residuals that constrain
the partial covering are much smaller (see Figure 6). We note that GIS 
calibration uncertainties almost certainly cannot make the low and high count
rate state contours in Figure 4 consistent with each other (as is to be expected
from Figure 6).  

We have examined whether a narrow 6.4 keV emission line from iron is
present in the high count rate state data by adding a Gaussian emission
line model. The addition of the line is not statistically significant, 
and the best fitting line normalization is zero. The line normalization
can be constrained to be less than $4.5\times 10^{-3}$ photons cm$^{-2}$ s$^{-1}$
with 90 per cent confidence ($\Delta\chi^2=2.71$). This corresponds to a
line equivalent width of less than 10 eV. While we find that the line
equivalent width is much smaller than for the low count rate state, 
we cannot formally prove that the line normalization is much smaller.  

We have analyzed other sets of data taken after 12000 s in Figure 1, and the
iron edge appears to be generally much weaker during this time than 
during the low count rate state. In particular, spectra created from data
taken during the two rises between 28000--36000 s in Figure 1 do not show
iron K residuals any stronger than in the bottom panel of Figure 6.

\subsubsection{Spectral ratios}

As discussed by Inoue (1989), the ratio of two spectra that differ only
in the fraction of thick partial covering imposed on them gives useful 
information about this partial covering. At X-ray energies of $<3$ keV, 
where photoelectric absorption blocks all X-rays
striking the partial covering material, the spectral ratio should approach
a constant value. This constant value gives the relative partial covering 
fractions in the two spectra. At higher X-ray energies, as photoelectric
absorption becomes progressively less important, the spectral ratio
should approach unity (this effect will not be observable in the \ASCA
band for columns significantly larger than 
$\sim \sigma_{\rm T}^{-1} \sim 1.5\times 10^{24}$ cm$^{-2}$). 
Exactly this behaviour was reported by Inoue (1989), and we have made 
ratios of our spectra to look for it.

We have made a ratio of 
the GIS2 spectrum from the `calm' set of low count rate state data to
the GIS2 spectrum from the bottom panel of Figure 6 
(`spectral ratio 1' or `SR1')
as well as a ratio of
the GIS2 spectrum from the HR3$>0.9$ set of low count rate state data to
the GIS2 spectrum from the bottom panel of Figure 6 
(`spectral ratio 2' or `SR2').
We show a plot of these ratios in Figure 7. SR2 starts to approach unity 
above 4 keV as its numerator's column is progressively penetrated by higher 
energy X-rays, and SR1 also starts to rise at the high energy end of the
\ASCA band (this behaviour will be able to be probed much better with
{\it RXTE\/}). The $<3$ keV behaviours of SR1 and SR2 are quite similar. However, 
these behaviours are both significantly different from that shown in Figure 4 of 
Inoue (1989). They resemble more closely the behaviour seen by Ikegami (1986). 
Neither spectral ratio is approximately constant below
3 keV as would be expected under the hypothesis of only a partial covering
fraction change between the low count rate state and the high count rate
state.  

Using the two blackbody continuum model with partial 
covering, we have also determined that the combination of 
a partial covering fraction change and a partial covering column change 
cannot, by itself, explain the difference between the low and high count 
rate state spectra. We have done this by jointly fitting the GIS spectra
for the `calm' low count rate state data and the high count rate state data.
In this fitting we tie all model parameters together other than those of the 
partial covering. The GIS2 and GIS3 blackbody normalizations are separately tied 
together for each pair of spectra. We include 2 per cent calibration
uncertainties. The resulting joint fit is statistically
unacceptable and can be rejected with greater than 99 per cent confidence. 
As expected from Figure 7, examination of the residuals shows that the low 
count rate state spectra are softer below $\approx 2$ keV than the high count 
rate state spectra. If we add a third blackbody that is unaffected by partial 
covering to our joint fitting, we can obtain an acceptable joint fit. The third
blackbody has a temperature of about 0.5 keV but is poorly constrained. 
 
The fits of the previous paragraph and their residuals could signify either 
(1) a true change of the blackbody parameters between the low and high
count rate states (as well as a partial covering change) or 
(2) the presence of an additional soft component that is unaffected by
the partial covering so that it can be seen during the low count rate state 
but is diluted away from observability during the high count rate state. 
As is clear from previous sections, our spectral fits alone cannot
formally falsify the first possibility. However, given the observed
spatial extent at low energy the additional soft component of the
second possibility can naturally be explained as time-delayed X-rays that 
have been dust scattered into the line of sight by 
the interstellar medium. For a source at the distance 
and position of Cir X-1, the time delay is $\sim 10$ days (see equation 20
of Mauche \& Gorenstein 1986), a significant fraction of the orbital 
period. Replacing the third blackbody of the previous paragraph by
a model consisting of the first two blackbodies multiplied by
$E^{-2}$ does give a fair fit to the data so indicating that
dust scattering is a plausible hypothesis. Our relatively 
short span of low count rate state data does not 
allow us to probe sensitively for temporal changes of the soft component due 
to the decay of the dust halo.

\begin{figure}
{\psfig{figure=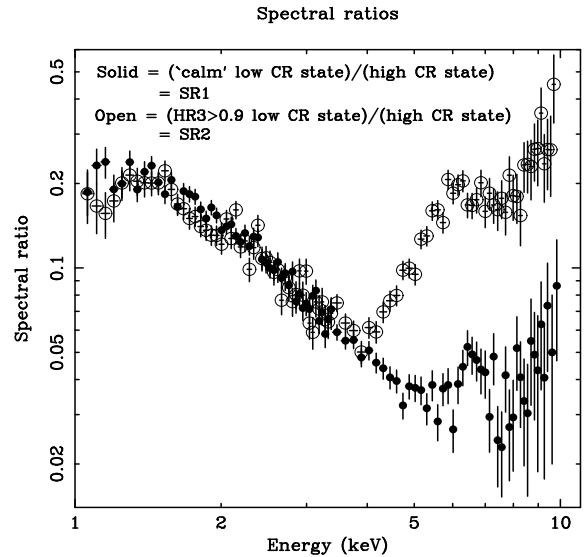,width=4.2 in,height=3.0 in,angle=270}}
\caption{GIS2 spectral ratios as labeled in the diagram 
(`CR' stands for `count rate' and `SR' stands for spectral ratio).}
\end{figure}
 
\section{Discussion and summary} 

\subsection{Partial covering: a direct flux plus electron scattered flux model}

Our observations strongly argue that partial covering is a crucial
determinant of the behaviour of Cir X-1. We clearly see the spectral
features expected from partial covering (Figure 3), and 
the observed continuum spectral 
variability is also best explained in terms of it (Figure 5). 
These two separate effects in conjunction are compelling. As mentioned 
in Section 3, a large change in the partial covering column 
without a corresponding change in the partial covering fraction 
(see Figure 4) is physically somewhat 
surprising. However, as we describe below, such a change could easily occur 
if we observed both direct, but attenuated, X-rays as well
as X-rays that are electron scattered around the attenuating matter
(this electron scattered component is local to Cir X-1 and totally distinct
from the dust scattered component of the previous section). 
Cir X-1 would then be like a Seyfert 2 galaxy with a Compton-thin and 
time-variable torus, and the fitted partial covering would really 
represent total covering plus a significant electron scattered component. 
Partial covering models have been used to fit Seyfert 2 X-ray spectra 
(e.g. Iwasawa et~al. 1994). As will become clear, the direct flux plus 
electron scattered flux scenario obviates the geometrically 
difficult requirement to partially cover an object as small as a
neutron star and its inner disc.

In this scenario, the direct X-rays would be completely blocked at low energies 
but would penetrate through the absorbing matter above $\sim$3--5 keV and 
thereby imprint the observed iron edge on the spectrum. The absorbing material
could be the outer bulge of an accretion disc if the disc is seen 
{\it nearly edge-on\/}, and the column changes could occur as the disc rotates
(dynamic changes of the partial coverer are possible as well). For a
disc outer radius of $\sim 1$ solar radius the disc rotation timescale is
such that this scenario appears plausible. Several
Seyfert 2 galaxies with Compton-thin tori also show large edges from
neutral iron (e.g. Awaki et~al. 1991). 

The electron scattering of X-rays could arise in an accretion 
disc corona. One potential prediction of models with a large amount of 
electron scattering is that the disc 
flux from Cir X-1 may be polarized over the expected interstellar value at 
wavelengths where the direct emission is highly absorbed (of course, one
will also have to take into account polarization dilution by the companion 
at wavelengths where this is relevant). The polarization fraction should
change with time as the absorbing column, and hence the direct contribution, 
changes. If scattered photons 
suffered multiple scatterings the polarization could be significantly 
reduced (indeed, if the Thomson depth of the scattering `mirror' were greatly
below unity the implied isotropic luminosity would be prohibitive). 
If the jets of Cir X-1 are perpendicular to its disc, one
might expect the polarization position angle to be perpendicular
to the jet position angle. 

From the previous paragraphs and observations of Seyfert 2 galaxies, one 
might expect to observe very large equivalent width ($\sim 1$ keV) iron K 
lines from Cir X-1 during the low count rate state. However, it is 
important to remember that some direct continuum X-rays are able to penetrate
through the obscuring matter at the energies of the iron K lines, and
these will reduce the line equivalent widths. 
Furthermore, the solid angle subtended by the accretion disc 
bulge need not be large, and this would reduce the equivalent width
of the iron line from neutral matter. Performing a calculation similar
to that shown in figure 1.11 of White, Nagase \& Parmar (1995), we
find that $\Delta\Omega/2\pi$ must be about 0.1 to explain the iron line
equivalent width, which is consistent with what one might 
expect from the outer bulge of an accretion disc. 
The ionized iron lines need not be 
strong if iron atoms in the scattering mirror are 
{\it fully stripped\/} of their electrons, and the 
large luminosity of Cir X-1 allows
iron atom stripping out to radii of $\sim 0.2$R$_\odot$ or more 
(we take the Thomson depth to be $\tau\approx 1$ and require
the ionization parameter to be $\xi\simgt 10^4$ erg cm s$^{-1}$;
see Kallman \& McCray 1982). Lower atomic number atoms would also
be fully stripped and thus would not produce lines. 

Finally, one might expect to see a broad iron line and/or smeared
edge due to X-ray reflection off the surface of the inner accretion
disc (e.g. Ross, Fabian \& Brandt 1996 and references therein).
No such features are clearly detected (see the bottom panel of
Figure 6). There are three possible reasons for the absence
of such features: (1) the magnetosphere or an instability disrupts 
the inner accretion disc so little disc reflection takes
place, (2) the disc inclination is high so that the reflection
component is relatively weak (see George \& Fabian 1991; note this is
what we have suggested above) and (3)      
electron scattering in the corona, if its temperature is $\simgt 1$ keV,
renders features indistinguishable by Doppler broadening 
(such Doppler broadening has only a small effect on the continuum 
shape).  

\subsection{Partial covering and the abrupt count rate change}

Our spectral fitting above suggests a strong reduction in the amount of
partial covering during the low to high count rate state transition  
(see Figure 1 and Figure 6). Furthermore, for our 
best fits, the partial covering corrected 
isotropic luminosities (hereafter `the true luminosities') 
in the low and high count rate states are quite similar 
(the absorption corrected fluxes are 
$3.8\times 10^{-8}$ erg cm$^{-2}$ s$^{-1}$ and 
$4.1\times 10^{-8}$ erg cm$^{-2}$ s$^{-1}$, respectively). 
Indeed, we find that for a large region of statistically acceptable chi-squared 
parameter space the true luminosity of 
Cir X-1 {\it need not\/} change during the observed transition.
It is more difficult to prove the stricter statement that the true
luminosity of Cir X-1 {\it does not\/} change, due to the fact that for very
thick partial covering the true luminosity and partial covering 
percentage are highly covariant quantities (cf. the contours in Figure 4).    
If the true luminosity of Cir X-1 does not change significantly between the
low and high count rate states, Cir X-1 would have been constantly accreting 
at or perhaps even above the Eddington rate throughout our observation,
despite the fact that its count rate varied by about an order of
magnitude. Furthermore, as discussed in section 3.2.3, the underlying
spectral shape {\it need not\/} change either provided there is a third 
soft component that is diluted to invisibility once the partial
covering is reduced (again it is hard to prove that the underlying spectral 
shape {\it does not\/} change). We note that, despite the lack of strong low 
energy absorption changes, much of the behaviour studied in 
Dower, Bradt and Morgan (1982) could 
also result from partial covering changes provided the partial
coverer is so thick that essentially all the X-rays striking 
it are entirely absorbed. Indeed, the famous figure 4 of 
Dower et~al. (1982) is highly suggestive of partial
covering by very thick matter. 

However, as discussed by Dower et~al. (1982) and other authors, the strong 
radio and infrared outbursts near zero phase cannot be obviously understood
solely in terms of partial covering changes. Furthermore, due to the
observed flux rises during the X-ray bursts, we know that the true 
luminosity of Cir X-1 cannot be the Eddington one at all times.
The recent flux rises to 2--3 Crab seen in \RXTE all-sky monitor
data also suggest true luminosity changes. The detailed 
situation is likely to be very complex, although our observations convincingly
demonstrate that some form of partial covering is a key element. 
It is perhaps possible that the obscuring matter is 
ejected from the vicinity of the neutron star at the time of zero phase, 
perhaps to make the radio nebula or jets. Our iron line measurements are
consistent with an ejection of matter although the
statistics are not good enough to prove that such an ejection must
occur (i.e. we cannot show that the line normalization must drop 
dramatically).     

Our observations do not appear to give direct insight into why Cir X-1 
accretes at such a remarkable rate, and we have little to add to the
theoretical speculation on this issue. It may well be difficult
to gain further insight into this matter until the orbital parameters, 
companion star spectral type, proper motion and neutron star magnetic field 
of Cir X-1 are better determined. Observations to measure these crucial
system parameters are needed.

\section*{Acknowledgments}

We gratefully acknowledge financial support from the United States 
National Science Foundation (WNB) and the Royal Society (ACF).
We thank K. Ebisawa, K. Iwasawa, K. Mukai, Ph. Podsiadlowski,
R. Ross, R. Shirey, Y. Ueda, N. White, L. Whitlock and R. Wijers 
for useful discussions. We thank the members of the \ASCA 
team who made these observations possible.

\bsp

\end{document}

%% file: table1.tex
\begin{table*}
\caption{Data screening criteria for the SIS and GIS detectors. Details on these screening criteria
may be found in section 5.2 of Day et~al. (1995b).}
\begin{tabular}{lll}
\hline
SIS screening criteria & GIS screening criteria & Description \\
\hline
{\tt SAA$=0$                   } & {\tt SAA$=0$          } & Only use data taken outside the South Atlantic Anomaly (boolean variable) \\
{\tt ELV$>5$                   } & {\tt ELV$>5$          } & Earth elevation angle for target (degrees) \\
{\tt COR$>4$                   } & {\tt COR$>4$          } & Minimum cut-off rigidity (GeV $c^{-1}$) \\
{\tt ANG\_DIST$>0.0$           } & {\tt ANG\_DIST$>0.0$  } & Angular distance of field of view from specified direction (degrees) \\   
{\tt ANG\_DIST$<0.01$          } & {\tt ANG\_DIST$<0.01$ } & Angular distance of field of view from specified direction (degrees) \\   
{\tt T\_SAA$<0$ or T\_SAA$>16$ } & ---                     & Avoid high background data just after SAA passage (seconds) \\
{\tt S0\_PIXL1$>0$             } & ---                     & Number of events detected by chip 1 above the event threshold (SIS0 only) \\
{\tt S0\_PIXL1$<400$           } & ---                     & Number of events detected by chip 1 above the event threshold (SIS0 only) \\
{\tt S1\_PIXL3$>0$             } & ---                     & Number of events detected by chip 3 above the event threshold (SIS1 only) \\
{\tt S1\_PIXL3$<400$           } & ---                     & Number of events detected by chip 3 above the event threshold (SIS1 only) \\
---                              & {\tt G2\_L1$>0.0$     } & Only use data inside the PHA discriminator range (GIS2 only) \\
---                              & {\tt G3\_L1$>0.0$     } & Only use data inside the PHA discriminator range (GIS3 only) \\
\hline
\end{tabular}
\end{table*}

%% file: table2.tex

\begin{table*}
\caption{Spectral fitting results.}

\begin{minipage}{7.0 in}
\noindent `CR' stands for `count rate', 
`PC' stands for `partial covering' and 
`BB' stands for blackbody. 

\vspace{0.06 in}

\noindent The fitting models used below are defined in Shafer et~al. (1991). 
When we fit blackbody models we are using the `bbodyrad' model, 
and when we fit partial covering models we are
using the `pcfabs' model. The blackbody model normalizations ($A_1$ and $A_2$) 
are formally related to the radii of the emission regions, although as commented
in the text the derived radii are likely to only be characteristic quantities. A
`bbodyrad'  normalization is formally equal to the square of the emission
region radius in kilometers divided by the square of the
distance to Cir X-1 in units of 10 kpc. 

\vspace{0.06 in}

\noindent The errors for all fits are quoted for 90 per cent confidence,
conservatively taking all free parameters to be of interest other than absolute
normalization (Lampton, Margon \& Bowyer 1976). 

\vspace{0.06 in}

\noindent Fluxes and normalizations are quoted for SIS0 when possible. When SIS 
data are not available due to telemetry saturation we quote fluxes and normalizations 
for GIS2. All fluxes and luminosities are for the 1--10 keV band.
When we compute unabsorbed fluxes we remove the effects of both interstellar
absorption and intrinsic partial covering.

\vspace{0.06 in}

\noindent `d.o.f.' stands for `degrees of freedom'.
The $P(\chi^2\mid\nu)$ chi-square probability function is defined in
section 6.2 of Press et~al. (1989). 

\vspace{0.06 in}

\end{minipage}

\begin{tabular}{lcccccc} 
\hline
\multicolumn{1}{l}{Fit} & 
\multicolumn{1}{c}{`Calm' low} & 
\multicolumn{1}{c}{`Calm' low} & 
\multicolumn{1}{c}{`Calm' low} &
\multicolumn{1}{c}{HR3$>0.9$ low} &
\multicolumn{1}{c}{High} &
\multicolumn{1}{c}{High} \cr
\multicolumn{1}{l}{parameter} & 
\multicolumn{1}{c}{CR state} & 
\multicolumn{1}{c}{CR state} & 
\multicolumn{1}{c}{CR state} &
\multicolumn{1}{c}{CR state} &
\multicolumn{1}{c}{CR state} &
\multicolumn{1}{c}{CR state} \cr
\multicolumn{1}{l}{name} & 
\multicolumn{1}{c}{No PC} & 
\multicolumn{1}{c}{PC on both BB} & 
\multicolumn{1}{c}{PC on hot BB} &
\multicolumn{1}{c}{PC on both BB} &
\multicolumn{1}{c}{No PC} &
\multicolumn{1}{c}{PC on both BB} \cr
\multicolumn{1}{l}{} & 
\multicolumn{1}{c}{(1)} & 
\multicolumn{1}{c}{(2)} & 
\multicolumn{1}{c}{(3)} &
\multicolumn{1}{c}{(4)} &
\multicolumn{1}{c}{(5)} &
\multicolumn{1}{c}{(6)} \cr
\hline                                       
Interstellar                            & \cr
N$_{\rm H}$/(10$^{22}$ cm$^{-2}$)       & $1.77^{+0.11}_{-0.11}$  
                                        & $1.79^{+0.15}_{-0.12}$ 
                                        & $1.80^{+0.16}_{-0.13}$ 
                                        & $1.82^{+0.23}_{-0.20}$ 
                                        & $1.82^{+0.08}_{-0.08}$ 
                                        & $1.79^{+0.11}_{-0.15}$ \cr\cr
$kT_1$ (keV)                            & $0.48^{+0.03}_{-0.03}$  
                                        & $0.46^{+0.03}_{-0.04}$ 
                                        & $0.46^{+0.03}_{-0.04}$ 
                                        & $0.46^{+0.05}_{-0.05}$ 
                                        & $0.59^{+0.03}_{-0.03}$ 
                                        & $0.59^{+0.07}_{-0.05}$ \cr\cr
$A_1$                                   & $3710^{+930}_{-710}$    
                                        & $(6.5^{+4.8}_{-4.2})\times 10^4$ 
                                        & $4150^{+1760}_{-1110}$ 
                                        & $(4.0^{+2.0}_{-1.2})\times 10^4$ 
                                        & $8950^{+2030}_{-1520}$ 
                                        & $(1.5^{+10.5}_{-0.7})\times 10^4$ \cr\cr
$kT_2$ (keV)                            & $1.49^{+0.15}_{-0.13}$  
                                        & $1.16^{+0.19}_{-0.15}$ 
                                        & $1.15^{+0.17}_{-0.15}$ 
                                        & $1.51^{+0.12}_{-0.13}$ 
                                        & $1.42^{+0.03}_{-0.03}$ 
                                        & $1.38^{+0.04}_{-0.05}$ \cr\cr
$A_2$                                   & $19^{+6}_{-6}$          
                                        & $775^{+980}_{-460}$ 
                                        & $810^{+2040}_{-600}$ 
                                        & $201^{+120}_{-68}$ 
                                        & $565^{+60}_{-63}$ 
                                        & $1080^{+5120}_{-730}$ \cr\cr
Partial covering                        & \cr
percentage                              & --- 
                                        & $93.8^{+4.8}_{-10.6}$ 
                                        & $93.8^{+4.7}_{-10.8}$ 
                                        & $89.6^{+5.0}_{-5.1}$ 
                                        & ---
                                        & $42^{+54}_{-31}$ \cr\cr
Partial covering                        & \cr
N$_{\rm H}$/(10$^{22}$ cm$^{-2}$)       & ---                     
                                        & $162^{+44}_{-41}$ 
                                        & $161^{+104}_{-58}$ 
                                        & $42^{+7}_{-6}$ 
                                        & ---
                                        & $123^{+109}_{-108}$ \cr\cr
Absorption uncorrected                  & \cr
flux/(10$^{-9}$ erg cm$^{-2}$ s$^{-1}$) & 1.5 
                                        & 1.5 
                                        & 1.5 
                                        & 3.5 
                                        & 24 
                                        & 24 \cr\cr
Absorption corrected                    & \cr
flux/(10$^{-9}$ erg cm$^{-2}$ s$^{-1}$) & 2.5 
                                        & 38 
                                        & 16 
                                        & 25 
                                        & 33 
                                        & 41 \cr\cr
$\chi^2$                                & 1001        
                                        & 931.6 
                                        & 931.3
                                        & 1075 
                                        & 1389 
                                        & 1364 \cr
d.o.f.                                  & 932 
                                        & 930 
                                        & 930 
                                        & 1079 
                                        & 1279 
                                        & 1277 \cr   
$\chi^2_\nu$                            & 1.074                   
                                        & 1.001 
                                        & 1.001 
                                        & 0.996 
                                        & 1.086 
                                        & 1.068 \cr\cr
$P(\chi^2\mid\nu)$ rejection            & \cr
probability (if significant)            & 94.2                    
                                        & --- 
                                        & --- 
                                        & --- 
                                        & 98.3 
                                        & 95.5 \cr
\hline
\end{tabular}
\end{table*}